\documentclass[copyright,creativecommons,sharealike]{eptcs}
\providecommand{\event}{ACL2-2013}
\usepackage{times}
\usepackage{graphicx}
\usepackage{setspace}
\usepackage{verbatim}
\usepackage{csquotes}
\usepackage{amsfonts}
\usepackage{fancyvrb}
\usepackage{chngpage}
\usepackage{pbox}

\newcommand{\graphicswidth}{.8\textwidth}
\newcommand{\twolinecell}[2][l]{%
  \begin{tabular}[#1]{@{}l@{}}#2\end{tabular}}

\title{Proof Pad: A New Development Environment for ACL2}
\author{Caleb Eggensperger
\institute{School of Computer Science\\
University of Oklahoma\\
Norman, Oklahoma}
\email{calebegg@gmail.com}
}
\def\titlerunning{Proof Pad}
\def\authorrunning{Caleb Eggensperger}
\begin{document}
\maketitle

\begin{abstract}
Most software development projects rely on Integrated Development Environments
(IDEs) based on the desktop paradigm, with an interactive, mouse-driven user
interface. The standard installation of ACL2, on the other hand, is designed to
work closely with Emacs. ACL2 experts, on the whole, like this mode of
operation, but students and other new programmers who have learned to program
with desktop IDEs often react negatively to the process of adapting to an
unfamiliar form of interaction.

This paper discusses Proof Pad, a new IDE for ACL2. Proof Pad is not the only
attempt to provide ACL2 IDEs catering to students and beginning programmers. The
ACL2 Sedan and DrACuLa systems arose from similar motivations. Proof Pad builds
on the work of those systems, while also taking into account the unique workflow
of the ACL2 theorem proving system.

The design of Proof Pad incorporated user feedback from the outset, and that
process continued through all stages of development. Feedback took the form of
direct observation of users interacting with the IDE as well as questionnaires
completed by users of Proof Pad and other ACL2 IDEs. The result is a streamlined
interface and fast, responsive system that supports using ACL2 as a programming
language and a theorem proving system. Proof Pad also provides a property-based
testing environment with random data generation and automated interpretation of
properties as ACL2 theorem definitions.
\end{abstract}

\section{Introduction and Prior Work}

\subsection{ACL2}
ACL2 (A Computational Logic for Applicative Common Lisp) is a lisp dialect and
theorem prover in the Boyer-Moore family of theorem provers. Functions and
theorems in ACL2 are stated using an applicative subset of Common Lisp, where
variables are read-only, and functions must be written using a recursive rather
than an iterative style. Because of this, user-stated theorems can be more
easily proven by the automated theorem prover from the collection of theorems
and lemmas available in the system (both built-in and user-defined). Using this
mechanism, it's possible to \enquote{steer} ACL2 to prove complex theorems with
minimal interaction \cite{car}. ACL2 has predominantly been used to model and
verify the hardware and sometimes software of critical computer systems.

Recently, Carl Eastlund  and Rex Page have used ACL2 as a pedagogic software
development tool
\cite{eastlund-acl2}\cite{page-se}\cite{page-proptest}\cite{hcw}. Page has been
using ACL2 and DrACuLa (an ACL2 IDE) in the University of Oklahoma's software
engineering course since 2003. ACL2 was brought into the course to aid the
students with good design and defect control in their projects.  The approach
has been successful, with students able to create projects of significant
complexity with a solid proof-based underpinning \cite{page-se}.

For this two-semester course, students were asked to approach a series of small
individual and team projects, leading up to a final, semester-long team
project. Projects begin with the creation of short functions and the
formalization of simple theorems, in order to develop students' ability to state
and prove properties of their code. The course culminates with a project of
2,000 to 3,000 lines, along with ACL2 properties for individual components,
documentation, unit tests, and integration tests
\cite{page-se}.

\subsection{Existing ACL2 development environments}

The development of Proof Pad was informed by existing tools for working with
ACL2. The most common of these is Emacs, which is suggested by the
documentation. Additionally, two attempts have been made at more user friendly
interfaces, both targeted at being easier to use for undergraduates: DrACuLa and
ACL2s.

I have had some prior experience working with ACL2 user interfaces in the form
of Try ACL2 (at \url{http://tryacl2.org}), an experimental website that I built
that gives a restricted, web-accessible REPL for ACL2 along with a simple
tutorial to help the user learn some of the basics of ACL2. Through this
project, I got some experience working with ACL2 programmatically along with
some feedback on this method of interaction from the programming community. Try
ACL2 is more of an introductory tool rather than a fully featured development
environment. Proof Pad builds on this work towards the goal of a more
comprehensive environment, bringing it in line with the tools mentioned above.

\subsubsection{DrACuLa}

DrACuLa \cite{dracula} is a plugin for DrRacket (formerly DrScheme) that acts as
an IDE for ACL2. It is primarily intended for classroom use. Because DrACuLa is
a DrRacket plugin, it has an excellent text editor for Lisp syntax. Its
parentheses matching and auto-indentation are indispensable. DrRacket also has a
history of pedagogic usage and, as such, is designed to be easy for students and
inexperienced programmers to pick up and use \cite{drracket}. DrACuLa also
extends ACL2 with support for a modular style of programming inspired by the
Scheme dialect that DrRacket uses. In this style, interfaces to modules are
separated from the implementation of the modules. The modules also have
contracts that implementations must uphold, which are mechanically verified by
ACL2 \cite{modular-acl2}.

In addition to supporting a sizable subset of ACL2's syntax and features,
DrACuLa extends ACL2 with a set of \enquote{teachpacks} --- short, easy-to-use
libraries with specific functionality. One notable example is DoubleCheck, a
QuickCheck-inspired library for running automated, randomized tests of ACL2
code. DoubleCheck provides both an accessible jumping-off point for students to
start thinking in terms of verifiable properties before going straight to
theorems, as well as a useful method of generating counterexamples for failed
theorems. The alternative, reading ACL2's output for a failed proof attempt, is
intimidating for many students \cite{doublecheck}.

However, a few aspects of DrACuLa pose problems for effective classroom use.
Installation of DrACuLa is a complex process that involves acquisition and
installation of three separate software components, interaction with the command
line, and some potential pitfalls that are difficult to recover from, especially
for beginning programmers. In order for this tool to be usable in courses at the
University of Oklahoma, I have compiled and maintained a lengthy document for
the installation process and investigated many potential ways to streamline the
process, all to no avail. Installing the software tools is a frequent
frustration in these classes.

Because DrACuLa executes ACL2 definitions in Scheme, only some of ACL2's
functionality is available. For example, arrays and macros are not supported,
and supporting them would likely involve a lot of work. This dual implementation
can also lead to bugs where certain functions can be used in DrACuLa, but are
rejected or incorrect in ACL2 (and vice versa). Additionally, this
dual-implementation can lead to error messages that are hard to follow or track
down in many cases, especially in the included teachpack libraries where there
are separate ACL2 and Racket implementations of each function.

\subsubsection{The ACL2 Sedan}
The ACL2 Sedan, or ACL2s, is an Eclipse plugin and set of additional features
for ACL2 that seek to make ACL2 easier to use. Using the metaphor of a sedan,
ACL2s intends to provide a simple, low-maintenance interface for ACL2 at the
cost of high-level performance and customization. ACL2s includes some extensions
of ACL2's functionality that add various useful features, including a method by
which it can automatically attempt to generate counter-examples directly from
theorems \cite{acl2s}.

ACL2s has an automatic counterexample generation tool that works by looking
directly at theorems to determine the types of data to try to bind to free
variables in the theorem body. This means that ACL2s doesn't require a different
syntax for specifying tests and data generators. This approach integrates more
directly with the existing proof process, as opposed to the DoubleCheck approach
of integrating proofs with an existing testing method \cite{acl2s-testing}.

However, ACL2s also has some drawbacks. The entire UI for ACL2s is placed in a
single toolbar consisting of nine buttons and a single top-level menu. This is
fine for Eclipse plugins that also re-use Eclipse's menu items and standard
functionality, but, for the most part, ACL2s does not. When looking at a file,
ACL2's status is shown only through colors, and no visual feedback is given for
errors.

ACL2s requires that Eclipse be installed in a non-standard path on Windows
systems. The initial placement of the REPL (which is treated as a type of file)
is in a separate tab, which the user must flip back and forth between to use,
instead of the docked window arrangement shown on the website. The default mode
that ACL2s users are put into is called "Bare Bones" mode, where common macros
such as addition, subtraction, and equality testing are undefined, with no clear
indication of how to change that.

Modern development environments frequently provide complex, language specific
syntax highlighting, that takes into account different categories and types of
keywords and built-in functionality, data types, and comments. ACL2s, however,
does not; syntax highlighting simply consists of one color for parentheses, one
for code, and one for comments. This makes typos difficult to spot and built-in
functionality difficult to recall (e.g. is it \verb+equal+ or \verb+equals+?).

\subsubsection{Emacs}
Professional users of ACL2 commonly use Emacs along with a plugin specifically
designed for ACL2. Emacs relies heavily on memorized keyboard commands and a
minimal, code-focused UI. This usage style is not practical for a classroom
environment where the students have learned programming on point-and-click IDEs,
and do not want to learn the numerous key commands necessary to use Emacs
proficiently. However, the emacs interface has several attributes that are
conducive to ACL2's unique workflow. I have tried to incorporate some of those
elements into Proof Pad.

\section{Design}

The design of Proof Pad was a carefully considered process. I sought to
fix some of the usability problems with previous attempts without
introducing new, worse problems in the process. I was also interested in
producing a polished, good-looking application that users would feel
comfortable and at home with. Careful design, fast prototyping and
iterating of design ideas, and user feedback were all core to this
process.

\subsection{Goals and Constraints}

\begin{table}[htbp]
\begin{adjustwidth}{1in}{1in}
\textbf{Describe three services an IDE should provide:}
\vspace{1ex}

\begin{tabular}{r|l}
    Count & Service \\
    \hline\\[-2ex]
    10 & \twolinecell{Syntax error messages pinpoint type of error\\ and
    location in source code} \\[2.5ex]
    7 & \twolinecell{Runtime error messages pinpoint type of error\\ and
    location in source code} \\[2.5ex]
    6 & Automatic indentation syntax highlighting, etc. \\
    5 & Syntax-error detection while typing \\
    5 & Runtime program stepper \\
\end{tabular}
\vspace{3ex}

\textbf{Describe three things you like about DrACuLa:}
\vspace{1ex}

\begin{tabular}{r|l}
    Count & Aspect \\
    \hline\\[-2ex]
    9 & Parentheses matcher during editing \\
    8 & DoubleCheck property based testing \\
    6 & ACL2's mechanized logic \\
    5 & Read-Eval-Print loop \\
    5 & \verb+check-expect+ tests \\
\end{tabular}

\vspace{3ex}

\textbf{Describe three things you don't like about DrACuLa:}
\vspace{1ex}

\begin{tabular}{r|l}
    Count & Aspect \\
    \hline\\[-2ex]
    10 & No runtime debugger or program stepper \\[1.5ex]
    8 & \twolinecell{Runtime error messages give little info about where\\ or
    what the error is (stack trace would help)}\\[2.5ex]
    7 & Syntax error messages fail to pinpoint error in source code \\
    5 & Slow start-up \\
    4 & Multi-tab editor hard to use \\
    4 & Poor documentation of intrinsic functions \\[1.5ex]
    4 & \twolinecell{Lack of auto-completion and display of function\\
    parameters while typing} \\
\end{tabular}

\end{adjustwidth}
\caption[DrACuLa questionnaire results]{Summary of results from the DrACuLa
    questionnaire administered in the University of Oklahoma's software
    engineering course in Fall of 2011. Only the top 5 most commonly mentioned
    items are shown (more if there's a tie)}
\label{dracula-table}
\end{table}

When designing Proof Pad, I started by documenting several goals and constraints
intended to make Proof Pad an effective, modern development environment. These
goals are based heavily on the results of a questionnaire completed by students
in the software engineering course at the University of Oklahoma. The
questionnaire posed three free-form questions, soliciting up to three answers to
each. The results of the questionnaire (see \autoref{dracula-table}) were
reviewed by Rex Page, Allen Smith, and myself as part of ongoing work in a
project investigating pedagogic use of ACL2. As part of this review, we
collected together responses we saw as duplicates, so that they could be
prioritized. The results revealed several areas of the existing platform,
DrACuLa, that seemed that they could be improved upon by taking a new approach.
Many of these elements could not be easily addressed in the DrRacket framework.

\subsubsection{User Interface}

Inasmuch as is possible, Proof Pad should have a feature set inspired by
industry standard IDEs such as Eclipse and Visual Studio. Professional and
pedagogic IDEs frequently include syntax highlighting, automatic indentation,
and bracket matching features to facilitate entering programs that are free of
syntax errors. These features are a must, especially to students who are used to
them in other environments.

Proof Pad should also encourage good use of ACL2 as a language. The Method is an
approach to proving theorems in ACL2 that was constructed and is advocated by
the developers of ACL2 \cite{car}. A document set up using The Method consists
of some proven theorems or lemmas, an invisible line separating the
proven/unproven portions, and then the unproven lemmas or theorems. Progressing
involves breaking down the first unproven theorem into lemmas to help steer ACL2
forward. Proof Pad can encourage users to follow this method of generating
proofs in at least two ways. First of all, it can display the proof line
prominently to help users keep track of where the proof process is. Another
aspect of The Method is the advice to read the proof transcript from the top to
see where ACL2 went wrong, instead of blindly trying to prove the last part of
the proof process that failed. For advanced users, this might involve a lot of
scrolling through irrelevant output, but for novice users attempting more simple
proofs, the ability to read and interpret a proof narrative is important, so
Proof Pad initially scrolls the proof output to the top.

\subsubsection{ACL2 Integration}

ACL2 can be frustrating to work with programmatically, but it is powerful and
fast. Despite some benefits of a secondary interpreter such as the one DrACuLa
uses (the possibility of step debugging and better memory management, for
instance), it seems to me that the disadvantages outweigh the advantages, and
that integrating directly with ACL2 is the better route.

The installation process is a user's first exposure to an application. As part
of being tightly coupled with ACL2, and in order to give a good first
impression, Proof Pad has a simple installation process that's comparable to the
installation of other applications, and, more specifically, does not take the
form of a separately installed plugin to another piece of software (such as
Eclipse or DrRacket), and does not require the user to procure resources, such
as the platform-specific ACL2 binary, from multiple sources.

\subsubsection{Performance}

Having a fast start-up time is an important part of the user's perception of the
speed and performance of the program overall, and tends to leave a more lasting
impression than other performance measures. In the development of Proof Pad, I
paid particular attention to this metric.

\subsubsection{Cross platform}
Having an application that both runs on the user's operating system
of choice and that also fits with their expectations for other applications on
that platform is important to giving the user a good impression of the software
and a solid framework for using it. Since ACL2 is already available on Windows,
Mac, and Linux, Proof Pad should (and does) provide comparable cross-platform
support. Additionally, even at the University of Oklahoma, which requires
Windows for certain classes, a survey of Applied Logic (one of the target
courses for Proof Pad) students' homework submissions determined that 36\% of
students use Macs, and 59\% use Windows (the remaining 5\% used Windows in a
computer lab).

As much as possible, the tool should stay consistent with the platform it is
running on, both graphically and interactively. As such, I worked towards
compliance with the Windows User Experience Interaction
Guidelines\footnote{\url{http://msdn.microsoft.com/en-us/library/windows/desktop/aa511258.aspx}},
the Mac OS X Human Interface
Guidelines\footnote{\url{http://developer.apple.com/library/mac/\#documentation/UserExperience/Conceptual/AppleHIGuidelines}},
and the GNOME Human Interface
Guidelines\footnote{\url{http://developer.gnome.org/hig-book/3.0/}}.

\subsubsection{User-focused process}

I sought feedback from students and other ACL2 users at all stages of
development, in both formal and informal contexts. User testing is an essential
part of creating a high quality, usable application. I want Proof Pad to be as
free of usability problems as I can make it, to allow users to focus on their
code, not the interface. I used a combination of one-on-one testing with
individual students completing tasks that I designed and more broad-scale
testing by using the tool in a classroom setting in two semester-long courses,
from which I collected more aggregate and quantitative data using a
questionnaire.

\subsection{User Interface Components}

\begin{figure}[htb]
\centering
\includegraphics[width=\graphicswidth]{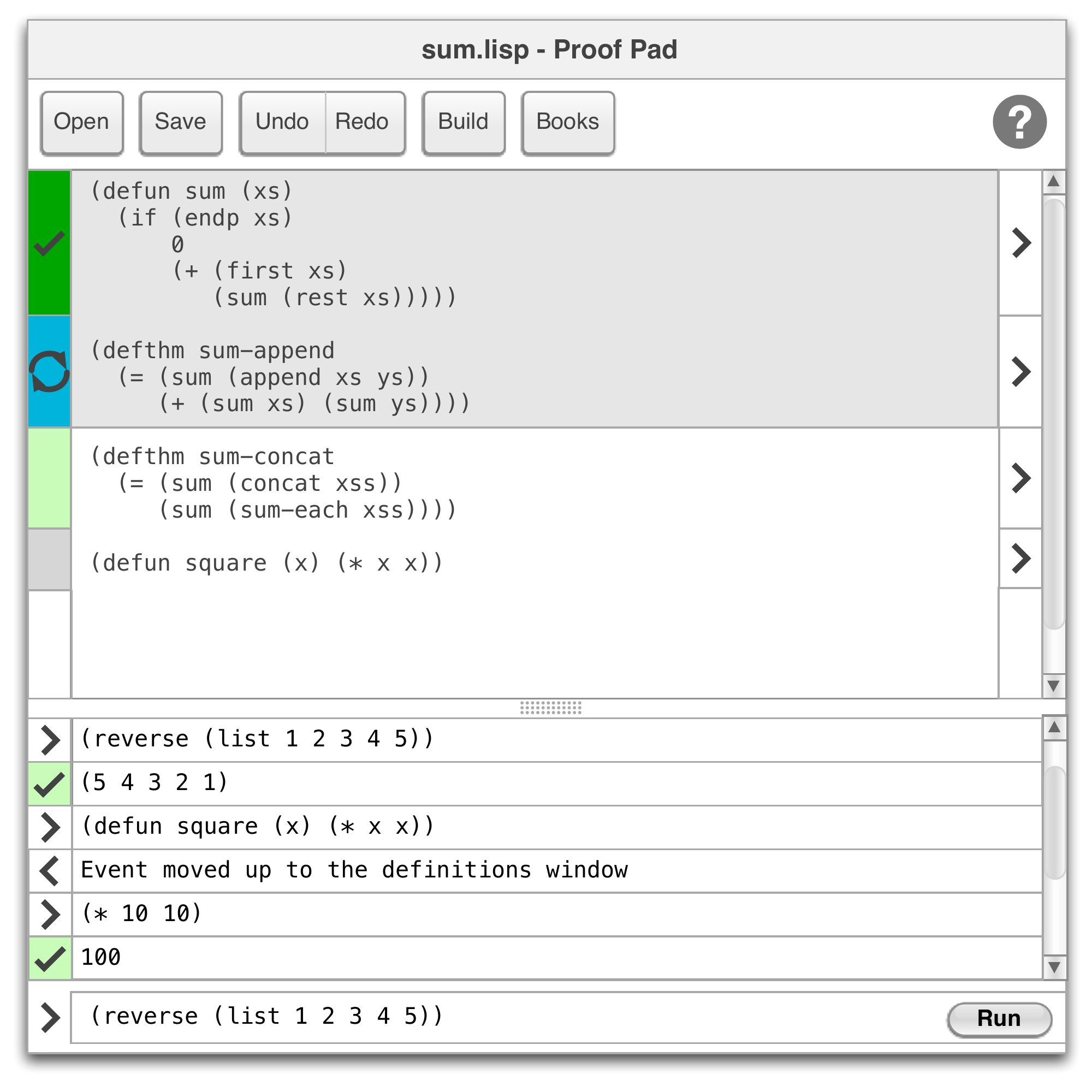}
\label{mainwindow}
\caption[Main proof pad window]{A mockup of the design of the main Proof Pad
window. A screenshot of the most recent version is shown in
\autoref{screenshot}.}
\end{figure}

The working area of Proof Pad is logically split into a few different areas. A
mockup of the main window is shown in \autoref{mainwindow}. The large area on
top where most of the code editing occurs is called the definitions area, since
the primary use for this part of the workflow is to define functions and
theorems. To the left of the definitions area is the proof bar, which displays
and allows manipulation of the state of ACL2 with respect to the definitions.
Below these items is the read-eval-print loop, an interactive console where
functions can be invoked to test or demonstrate their functionality.

\subsubsection{Definitions Area}
The definitions area is where files are edited. Functions, properties, and
theorems are defined in this area. The space is visually divided into areas that
denote the state of ACL2: the Proof Bar (see \autoref{proofbar-section}), the
actual syntax-highlighted text area, and a bar to the right that allows the user
to view the output of ACL2 for specific items.

The two bars to the left and right visually segment the definitions area based
on the top-level events or function calls in the definitions area. This
segmentation is further emphasized if part of the definitions have been admitted
to ACL2, which renders them with a grey background to indicate that they are not
editable and with a dark line below them to separate admitted and unadmitted
items.

The syntax highlighting has been carefully tailored to ACL2. For example, ACL2
makes a fundamental distinction between regular functions (such as
\verb+equals+, \verb+cons+, and \verb+append+), and events, which modify the
ACL2 state (such as \verb+defun+ and \verb+defthm+). Proof Pad emphasises this
difference by displaying the two types in different colors.

\subsubsection{Proof Bar} \label{proofbar-section}

\begin{figure}[htb]
\centering
\includegraphics[width=.9\textwidth]{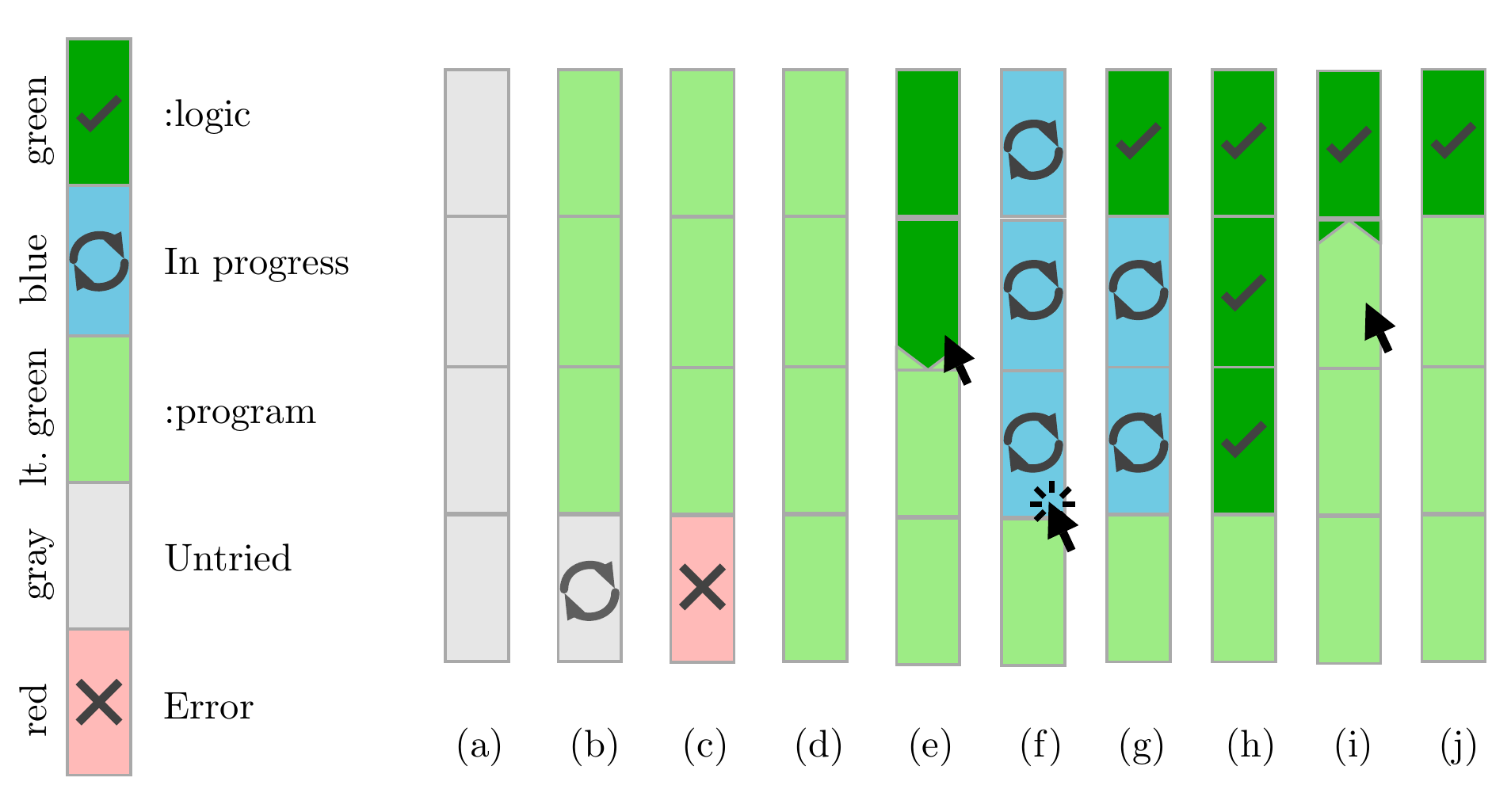}
\caption[Proof bar interactions]{Diagram showing how the proof bar responds to
    user interaction in some routine scenarios as a series of state transitions.
(a) Initial state.
(b) Three expressions automatically admitted.
(c) Fourth expression has an error.
(d) The error has been corrected.
(e) User hovers over the bar.
(f) User clicks the bar. Admission begins.
(g) One expressions has been admitted successfully to the logic.
(h) All three requested expressions admitted successfully.
(i) User hovers over second entry.
(j) User has unadmitted two expressions.}
\label{proofbar}
\end{figure}

Proof Pad introduces a user interface element for displaying and maintaining
the status of the current document with relation to ACL2. Typical ACL2 workflow
involves typing in a definition or theorem, attempting to admit it to ACL2, and
either responding to errors or continuing with the next definition.
Additionally, users occasionally need to return to a previously admitted
expression and modify it to accommodate a case or error that they had not
previously seen. \autoref{proofbar} shows some of the interactions the proof bar
supports.

In order to accommodate these use cases, the proof bar responds to clicks in one
of two ways:
\begin{enumerate}
\item If the click is to the left of an unadmitted expression, the proof bar
    queues for admission all of the unadmitted top-level expressions down
    through (and including) the selected expression for admission. As ACL2
    admits each one, the status changes from in progress to proven or failed.
\item If the click is to the left of an admitted expression, the proof bar
    undoes all of the admissions that changed the global state up through and
    including the selected expression. Undoing is done through the undo feature
    of ACL2, which has a fast response time.
\end{enumerate}
Hovering over the proof bar previews the action that will be taken if the user
clicks.

At all times, the proof bar indicates the current status of each top-level
expression in the definitions area. There are five different statuses that are
represented using different colors and symbols.

\subsubsection{Read-eval-print loop}
The read-eval-print loop (REPL) is a common feature of Lisp-like languages.
Formulas are typed at a prompt and executed in the context of the current set of
definitions. A REPL is a popular and effective part of other pedagogic IDEs,
such as DrRacket \cite{drracket} and DrJava \cite{drjava}.

One issue with the traditional REPL that the developers of DrRacket discovered
is that novice users often have trouble keeping track of stateful changes to the
environment made using the REPL. For instance, a student might fix a bug by
redefining a function in the REPL, but then forget to incorporate that change
back into the definitions area. This can lead to bugs that were thought to be
fixed returning later \cite{drracket}. Proof Pad addresses this problem by
keeping definitions consistent with updates to the REPL. All functions in ACL2
are divided into two groups: events, such as \verb+defun+ or \verb+defthm+,
which modify the global state; and functions, such as \verb+max+ or
\verb+first+, which do not modify the state of the system, but just return their
result. Proof Pad automatically moves event formulas to the definitions area
without passing them through ACL2, sidestepping the problem of state changes in
the REPL.

Proof Pad's REPL has some unique enhancements over typical REPLs. For one, Proof
Pad's REPL, like other parts of the UI, summarizes ACL2 responses to events,
while still allowing the user to click a disclosure arrow in the UI to view the
full ACL2 output.

\subsubsection{Results window}

\begin{figure}[htbp]
\begin{singlespace}
\begin{Verbatim}[frame=single,fontsize=\footnotesize]
ACL2 Warning [Compiled file] in ( INCLUDE-BOOK "book" ...):  Unable to load
compiled file for book
  <book path>
because that book is not certified.  See :DOC include-book.  No load was in
progress for any parent book.

ACL2 Error in ( INCLUDE-BOOK "book" ...):  There is no file named
"/Users/calebegg/Code/book.lisp" that can be opened for input.

Summary
Form:  ( INCLUDE-BOOK "book" ...)
Rules: NIL
Warnings:  Compiled file
Time:  0.00 seconds (prove: 0.00, print: 0.00, other: 0.00)

ACL2 Error in ( INCLUDE-BOOK "book" ...):  See :DOC failure.

******** FAILED ********
\end{Verbatim}
\end{singlespace}
\caption[ACL2 error message]{Full text of the error message shown when an
    included book can't be found. Collected from ACL2 version 4.3.}
\label{include-book-error}
\end{figure}

\begin{figure}[htb]
\centering
\includegraphics[width=\graphicswidth]{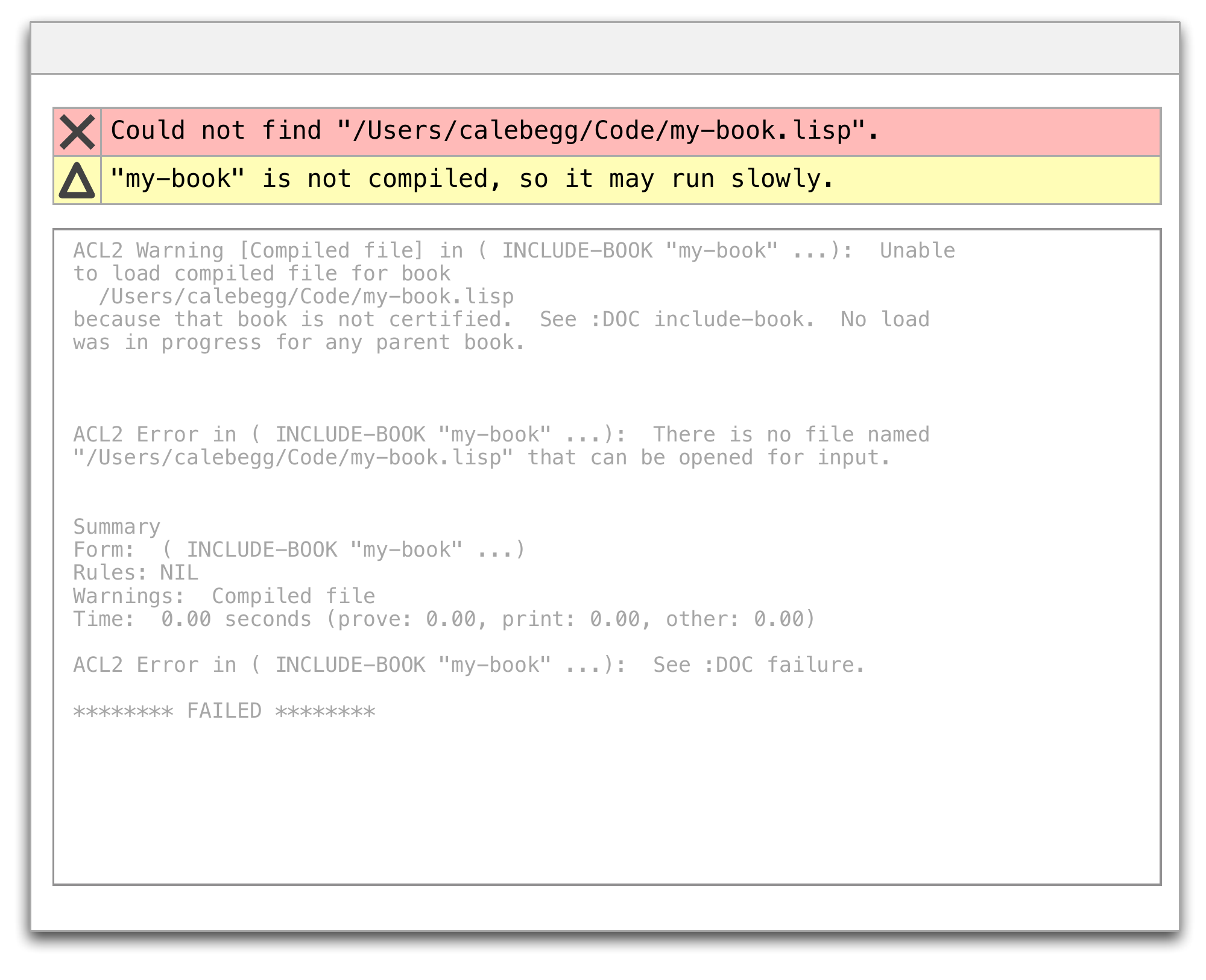}
\caption[Mockup of a warning message]{Mockup showing the results of trying to
import a non-existent book; note that the warning is de-emphasized in favor of
the more important error, but that the whole error message is also shown for the
user's perusal.}
\label{warning}
\end{figure}

When Proof Pad encounters an error, or when the user clicks on one of the
disclosure arrows either in the REPL or in the definitions area, the results
window appears to the right of the main window. The format of the results window
depends on the type of data being displayed. In most instances, it consists of
a color-coded summary of error, warning, and success messages that Proof Pad has
detected in ACL2's output, followed by a text area containing the raw ACL2
output.

A major part of this effort is the task of creating clear, concise summaries of
common or frustrating ACL2 error messages. ACL2's output tends to be verbose and
to suggest solution strategies that don't make sense for novice users. For
example, consider the ACL2 error message shown in \autoref{include-book-error}.
The first part of this message (where many students get stuck) suggests that the
user needs to certify the book, which leads the user to research the complex
certification feature of ACL2, even though, as is clear if the user keeps
reading, the real problem is that the book simply could not be found. Proof Pad
simplifies this to the error shown in \autoref{warning}. The warning and error
switch places (as they are sorted by importance), and the warning is
de-emphasized with color coding and a different icon.

\section{Implementation}

In total, Proof Pad comprises approximately 11,000 lines of Java code developed
over a period of 13 months. \autoref{timeline} shows a timeline for the project.
The project has expanded significantly as it progressed. The full source code
can be found at \url{http://github.com/calebegg/proof-pad}, and is available
under the GPLv3 license.

\begin{figure}[htb]
    \centering
    \includegraphics[width=4.76 in]{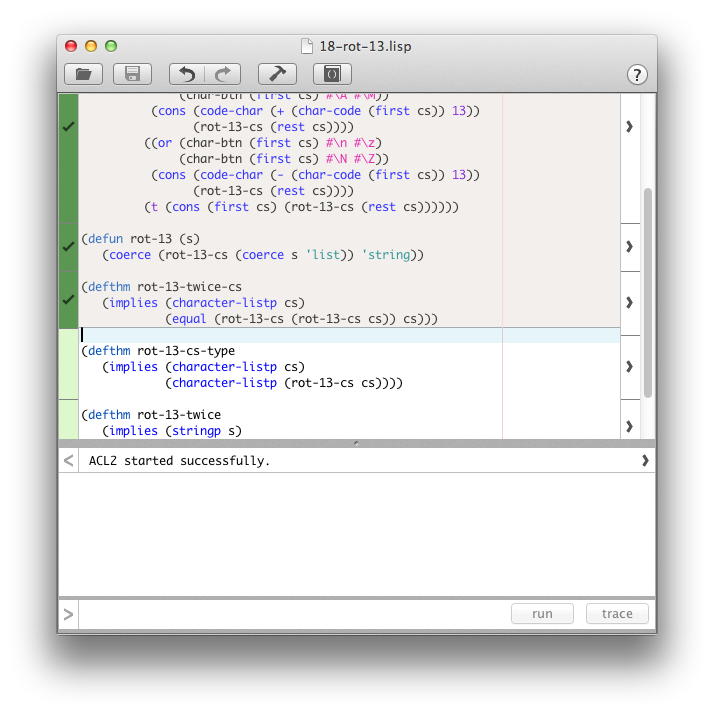}
    \caption[Main proof pad window]{The main Proof Pad window, showing a file
        that is being actively worked on.}
    \label{screenshot}
\end{figure}

\begin{figure}[ht]
    \centering
\begin{tabular}{r|ll}
    \textbf{March 2012}    & Began work on prototype & 7,400 lines of code \\
    \textbf{April 27 2012} & First user evaluation & 8,800 \\ [1.25ex] 
    \textbf{Fall 2012}     & Trial use in logic course & 9,800 \\
    \textbf{December 2012} & Survey of students in logic course \\ [1.25ex]
    \textbf{Spring 2013}   & Trial use in logic course & 10,900\\
    \textbf{March 2013}    & Second evaluation \\
\end{tabular}
\label{timeline}
\caption[Timeline of progress on Proof Pad]{Timeline showing some of the
major events in the development of Proof Pad.}
\end{figure}

\subsection{Code editor and syntax highlighting}
Proof Pad uses the open-source Java code editor \verb+RSyntaxTextArea+ for the
definitions pane and for the input to the REPL. \verb+RSyntaxTextArea+ makes
certain familiar parts of code editing, such as syntax highlighting, easier to
implement. I selected \verb+RSyntaxTextArea+ primarily for its speed and
simplicity. The syntax highlighting parser works with a JFlex lexical analyzer
generator. JFlex generates a minimal DFA from a regular expression based lexer
definition so that the lexer code it generates is performant. As part of
tailoring the syntax highlighting to ACL2, The JFlex lexer provided for Lisp
needed to be heavily modified to make it aware of ACL2's keywords and some
specific aspects of ACL2's primitive data types.

In addition to the syntax highlighting lexer, Proof Pad also has a slower,
simple parser that runs after a few seconds of inactivity and scans the user's
code, using the lexer's tokens, looking for several types of simple syntax
errors, such as undefined functions or variables, too many or too few arguments
for a built-in function or macro, or incorrect syntax for a built-in macro like
\verb+let+ or \verb+defthm+.

Proof Pad's auto-indentation feature watches for the user to enter a newline and
determines the proper level of indentation using a full trace of the previous
indent amounts and parentheses depth, taking into account specific indentation
patterns for certain built-in macros. The user can also re-indent sections of
code.

\subsection{Wrapping ACL2}
ACL2 is a complex tool with no public API. As such, it needs to be wrapped and
its output parsed in order for the UI to give the user feedback on its progress
and status. I used a separate Java thread and the Java standard library
\verb+ProcessBuilder+ API to call and keep up with the progress of ACL2. The
output is parsed and scanned for error or success messages, as well as for the
prompt sequence (\verb+ACL2 >+) to determine when the system is finished
admitting an expression. Additionally, since some types of functions and macros
can create multiple prompts, commands are sent to ACL2 to print identifying
strings that can then be scanned for in the output.

\subsection{DoubleCheck}
I ported part of DrACuLa's DoubleCheck library to pure ACL2, so that DoubleCheck
properties could be run in Proof Pad. This necessitated some changes to the
system resulting from limitations in ACL2. Probably most importantly,
DoubleCheck relies on second order functions to allow users to define new random
data generators. Since Proof Pad uses native ACL2 to interpret user code, Proof
Pad cannot support user-defined generators. At this point, Proof Pad implements
only a subset of the random data generators in Dracula's DoubleCheck system.
This limits the usefulness of property based testing in large software
development projects, but succeeds in allowing students to get some experience
in property-based testing.

\section{Evaluation}
\subsection{First user evaluation}
Participants in the first Proof Pad evaluation came from The University of
Oklahoma's software engineering course. This is the second semester of the
capstone course for Computer Science, so many of the participants were seniors.
Furthermore, because the course is taught using ACL2 in the DrACuLa environment,
the participants already had a moderate amount of experience using ACL2, but
with a different IDE. This made it possible to ask students participating in the
evaluation to perform sophisticated evaluation tasks.

Some studies of usability testing have found that for qualitative usability
studies like this one, small numbers of participants (four or five) are nearly
as effective at finding most usability problems in an application as large
numbers \cite{subjects}. Five students participated in this initial evaluation.

The evaluation consisted of four tasks, each taking at most five minutes,
followed by ten minutes allotted to discuss overall impressions of the tool,
for a total of 30 minutes per student. Evaluations were performed individually
with me on a Macbook Air, using either a Mac-like or Windows-like (with
Windows-style keybindings and menus) build of Proof Pad. All students chose the
Windows-style build.

For the first two tasks, the task was presented verbally alongside a Proof Pad
window populated with some code. In the first task, the participant was asked to
admit and run a provided function. In the second, a file that has three errors
was shown, and the participant was asked to identify and fix any of them that
they could, and then run the function.

For the third and fourth task, instructions were provided both verbally and on
screen, along with an empty Proof Pad window for the student to work in. The
first of these tasks asked them to write a function that computes the product of
a list; one example is provided. The second task asks them to write and
admit/run a theorem or property (DoubleCheck test) to demonstrate that
\verb+(prod (append xs ys))+ $=$ \verb+(* (prod xs) (prod ys))+, using either
their \verb+prod+ function from the previous task, or a provided one if they did
not complete the task.

From this first evaluation, I gained much valuable feedback. Nearly all
participants had trouble discovering the use and purpose of the Proof Bar.
Several participants expressed a desire for some of the standard features of
IDEs they've used before, such as line numbers, code folding, automatic
insertion of parentheses, automatic saving, options to automatically fix
highlighted errors, etc. In addition to this higher-level feedback, I also made
several small usability changes to the application based on more minor problems
experienced by only one or two participants.

\subsection{Use in logic courses (Fall 2012 and Spring 2013)}
\begin{figure}[htb]
\centering
\includegraphics[width=.6\textwidth]{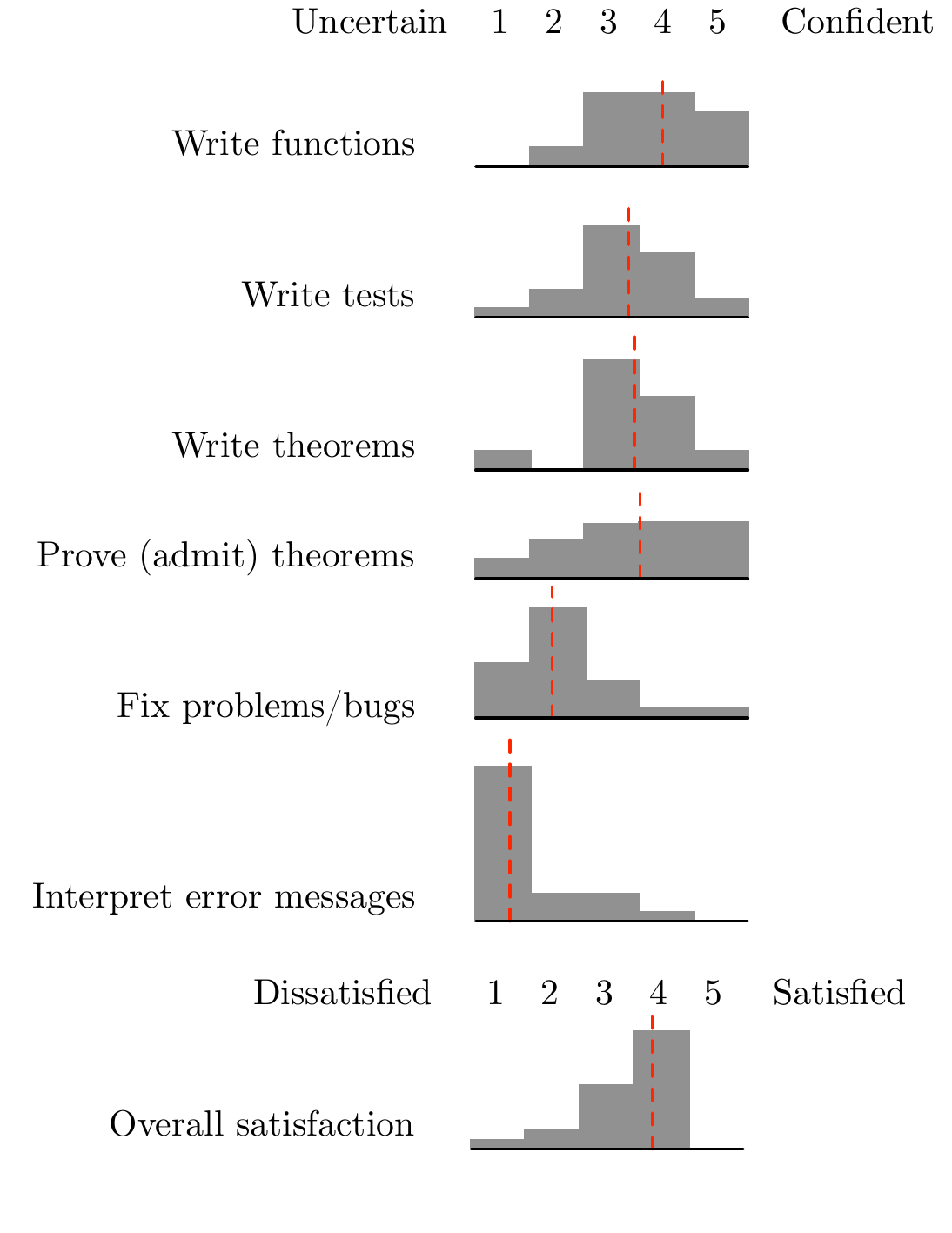}
\caption[Quantitative survey results]{Graphical summary of the quantitative data
collected from the one-page survey given to Applied Logic students at the end of
the Fall 2012 semester.}
\label{survey-results}
\end{figure}

Proof Pad was used as the primary software tool in the Fall 2012 section of
Applied Logic and the Spring 2013 section of How Computers Work\cite{hcw}, a
cross-listed honors course, at the University of Oklahoma. I provided support
for the software throughout these two semesters, and the students completed
several assignments using it. At the end of the semester, they were given an
anonymous survey where they were asked to rate their confidence using certain
parts of Proof Pad on a five-point scale, to supply (up to) three good things,
bad things, and desired features for Proof Pad, and, finally, to rate their
overall experience with Proof Pad on a five point scale. The results of the
quantitative sections for the fall 2012 semester are summarized in
\autoref{survey-results}. For the three qualitative sections, I collected the
data in a spreadsheet, combining entries that I saw as duplicates. The top three
items from each category are summarized in \autoref{results-table}.

\begin{table}[htb]
\begin{adjustwidth}{1in}{1in}
\textbf{List up to three things you liked, found easy to do, or thought were
useful about Proof Pad:}
\vspace{1ex}

\begin{tabular}{r|l}
    Count & Feature \\
    \hline\\[-2ex]
    5 & Simple/elegant interface \\
    5 & Ability to prove theorems \\
    3 & Speed \\
    3 & Visual feedback/proof bar \\
\end{tabular}
\vspace{3ex}

\textbf{List up to three things you did not like or found frustrating about
Proof Pad:}
\vspace{1ex}

\begin{tabular}{r|l}
    Count & Aspect \\
    \hline\\[-2ex]
    18 & Poor/difficult to understand error messages \\
     7 & Compatibility with individual computers \\
     4 & Resizing of the window is buggy \\
     3 & DoubleCheck syntax differs from textbook \\
\end{tabular}
\vspace{3ex}

\textbf{List up to three features you wish Proof Pad had:}
\vspace{1ex}

\begin{tabular}{r|l}
    Count & Feature \\
    \hline\\[-2ex]
    9 & Function documentation \\
    6 & Mark lines with errors \\
    4 & User manual/application help \\
    4 & Better error messages \\
    4 & Line numbers \\
\end{tabular}
\end{adjustwidth}
\caption[Qualitative survey results]{A summary of the qualitative results of the
survey. Responses were collected in a spreadsheet, collated based on their
similarity, and then sorted. Items with more than two responders are shown for
each of the three qualitative questions.}
\label{results-table}
\end{table}

\subsection{Second evaluation}

Another evaluation of Proof Pad is planned for the end of the spring of 2013
semester. I plan to recruit participants from the University of Oklahoma's
second-level introductory programming course, give them an IDE-agnostic training
session for the basics of ACL2, and then have them perform tasks one-on-one
using either Proof Pad or DrACuLa (randomly selected).

\section{Conclusion}

Proof Pad is available now at \url{http://proofpad.org/}. It has been used in
two classes at the University of Oklahoma, and will continue to see use there.
By putting pedagogic goals and student expectations at the forefront of all my
decisions, I hope to have developed a useful tool for teachers and students
alike to take advantage of ACL2 in course work.

\section{Future work}

\subsection{\texttt{.proofpad} files} \label{proofpad-files}

\begin{figure}[htb]
\centering
\includegraphics[width=.7\textwidth]{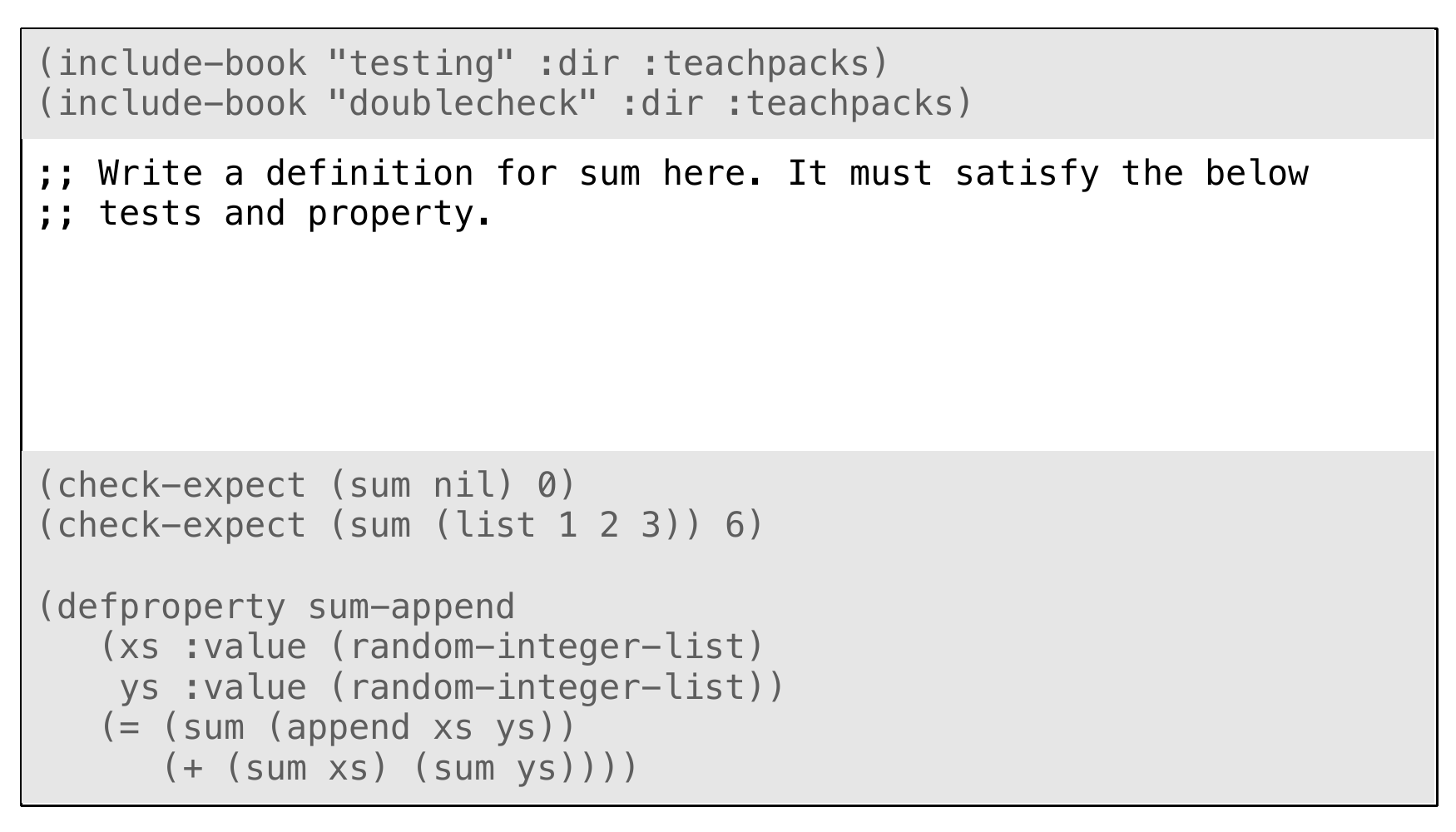}
\caption[Proof Pad files]{A mockup of the definitions area of Proof Pad, showing
an open \texttt{.proofpad} file.}
\label{proofpad-file}
\end{figure}

Since use of Proof Pad in the classroom is my primary goal, I think it's
important to take into account how ACL2 is used and how other pedagogic IDEs
work, and to try to design Proof Pad to be as easy to use in the classroom as
possible. In pursuit of this goal, and in line with some of the existing uses of
ACL2 in the classroom, I have a planned feature to make distributing code for
projects and giving students direct feedback on their progress possible.

Instructors will be able to create a special type of \verb+.proofpad+ file that,
when opened in Proof Pad, is divided into some read-write regions and some
read-only regions. Using this mechanism, an instructor can both provide existing
code (to supplement ACL2's limited standard functionality, for instance) and set
up tests or theorems that they expect the student to satisfy. An example is
shown in \autoref{proofpad-file}; in this example, I made use of DrACuLa's
\verb+testing+ teachpack, which includes a function, \texttt{(check-expect left
right)}, which passes if \verb+left+ $=$ \verb+right+, but fails and issues an
error message otherwise.

The student's goal for this file is to modify the read-write (light background)
portion of the file to make the following two tests and one property pass. In
this case, they can see that they need to write a function, \verb+sum+, that
returns either 0 for an empty list, or 6 for the list containing 1, 2, and 3.

Through this mechanism, the instructor can easily work up in difficulty from
simple tasks like modifying existing functions to cause failing tests to pass,
all the way to advanced tasks like steering ACL2 to solve a complex theorem by
adding lemmas to the logical world.

\section{Acknowledgements}

Rex Page is my Master's thesis advisor, under whom I've done this work. His
support and feedback have been invaluable. Peter Reid provided feedback and
several patches for issues he encountered early in the project. Matthew Kaney
created the icon for Proof Pad, selected the color scheme for the proof bar, and
gave feedback on other visual elements of the design. An early discussion about
Proof Pad with Ruben Gamboa led to the idea for the \verb+.proofpad+ files.
Proof Pad uses some open source components; in particular, the
previously-discussed \verb+RSyntaxTextArea+ (written and maintained by Robert
Futrell), and, of course, ACL2 itself. The compiled and certified binaries of
ACL2 that Proof Pad uses were created and provided by the ACL2s project.
Parts of DoubleCheck and several other DrACuLa libraries, written by Carl
Eastlund and others, are included with Proof Pad for compatibility with files
that use these libraries.

\bibliographystyle{eptcs}
\bibliography{thesis}

\begin{thebibliography}{10}
\providecommand{\bibitemdeclare}[2]{}
\providecommand{\surnamestart}{}
\providecommand{\surnameend}{}
\providecommand{\urlprefix}{Available at }
\providecommand{\url}[1]{\texttt{#1}}
\providecommand{\href}[2]{\texttt{#2}}
\providecommand{\urlalt}[2]{\href{#1}{#2}}
\providecommand{\doi}[1]{doi:\urlalt{http://dx.doi.org/#1}{#1}}
\providecommand{\bibinfo}[2]{#2}

\bibitemdeclare{inproceedings}{drjava}
\bibitem{drjava}
\bibinfo{author}{Eric \surnamestart Allen\surnameend}, \bibinfo{author}{Robert
  \surnamestart Cartwright\surnameend} \& \bibinfo{author}{Brian \surnamestart
  Stoler\surnameend} (\bibinfo{year}{2002}): \emph{\bibinfo{title}{DrJava: a
  lightweight pedagogic environment for Java}}.
\newblock In: {\sl \bibinfo{booktitle}{Proceedings of the 33rd SIGCSE technical
  symposium on Computer science education}}, \bibinfo{series}{SIGCSE '02},
  \bibinfo{publisher}{ACM}, \bibinfo{address}{New York, NY, USA}, pp.
  \bibinfo{pages}{137--141}, \doi{10.1145/563340.563395}.

\bibitemdeclare{inproceedings}{acl2s-testing}
\bibitem{acl2s-testing}
\bibinfo{author}{Harsh~Raju \surnamestart Chamarthi\surnameend},
  \bibinfo{author}{Peter~C. \surnamestart Dillinger\surnameend},
  \bibinfo{author}{Matt \surnamestart Kaufmann\surnameend} \&
  \bibinfo{author}{Panagiotis \surnamestart Manolios\surnameend}
  (\bibinfo{year}{2011}): \emph{\bibinfo{title}{Integrating Testing and
  Interactive Theorem Proving}}.
\newblock In \bibinfo{editor}{David \surnamestart Hardin\surnameend} \&
  \bibinfo{editor}{Julien \surnamestart Schmaltz\surnameend}, editors: {\sl
  \bibinfo{booktitle}{{\rm Proceedings 10th International Workshop} on the ACL2
  Theorem Prover and its Applications, {\rm Austin, Texas, USA, November 3-4,
  2011}}}, {\sl \bibinfo{series}{Electronic Proceedings in Theoretical Computer
  Science}}~\bibinfo{volume}{70}, \bibinfo{publisher}{Open Publishing
  Association}, pp. \bibinfo{pages}{4--19}, \doi{10.4204/EPTCS.70.1}.

\bibitemdeclare{inproceedings}{acl2s}
\bibitem{acl2s}
\bibinfo{author}{Peter~C. \surnamestart Dillinger\surnameend},
  \bibinfo{author}{Panagiotis \surnamestart Manolios\surnameend},
  \bibinfo{author}{Daron \surnamestart Vroon\surnameend} \&
  \bibinfo{author}{J.~Strother \surnamestart Moore\surnameend}
  (\bibinfo{year}{2007}): \emph{\bibinfo{title}{{ACL2s}: ``{T}he {ACL2}
  {S}edan''}}.
\newblock In: {\sl \bibinfo{booktitle}{Proceedings of the 7th Workshop on User
  Interfaces for Theorem Provers (UITP 2006)}}, {\sl
  \bibinfo{series}{Electronic Notes in Theoretical Computer Science}}
  \bibinfo{volume}{174}, pp. \bibinfo{pages}{3 -- 18},
  \doi{10.1016/j.entcs.2006.09.018}.

\bibitemdeclare{article}{modular-acl2}
\bibitem{modular-acl2}
\bibinfo{author}{C.~\surnamestart Eastlund\surnameend} \&
  \bibinfo{author}{M.~\surnamestart Felleisen\surnameend}
  (\bibinfo{year}{2009}): \emph{\bibinfo{title}{Toward a practical module
  system for ACL2}}.
\newblock {\sl \bibinfo{journal}{Practical Aspects of Declarative Languages}},
  pp. \bibinfo{pages}{46--60}, \doi{10.1007/978-3-540-92995-6\_4}.

\bibitemdeclare{inproceedings}{doublecheck}
\bibitem{doublecheck}
\bibinfo{author}{Carl \surnamestart Eastlund\surnameend}
  (\bibinfo{year}{2009}): \emph{\bibinfo{title}{{DoubleCheck} your theorems}}.
\newblock In: {\sl \bibinfo{booktitle}{Proceedings of the Eighth International
  Workshop on the ACL2 Theorem Prover and its Applications}},
  \bibinfo{series}{ACL2 '09}, \bibinfo{publisher}{ACM}, \bibinfo{address}{New
  York, NY, USA}, pp. \bibinfo{pages}{42--46}, \doi{10.1145/1637837.1637844}.

\bibitemdeclare{inproceedings}{eastlund-acl2}
\bibitem{eastlund-acl2}
\bibinfo{author}{Carl \surnamestart Eastlund\surnameend}, \bibinfo{author}{Dale
  \surnamestart Vaillancourt\surnameend} \& \bibinfo{author}{Matthias
  \surnamestart Felleisen\surnameend} (\bibinfo{year}{2007}):
  \emph{\bibinfo{title}{{ACL2} for Freshmen: First Experiences}}.
\newblock In: {\sl \bibinfo{booktitle}{ACL2 ’07: Proceedings of the Sixth
  International Workshop on the ACL2 Theorem Prover and its Applications}},
  \bibinfo{publisher}{ACM Press}, pp. \bibinfo{pages}{200–--211}.

\bibitemdeclare{incollection}{drracket}
\bibitem{drracket}
\bibinfo{author}{Robert \surnamestart Findler\surnameend},
  \bibinfo{author}{Cormac \surnamestart Flanagan\surnameend},
  \bibinfo{author}{Matthew \surnamestart Flatt\surnameend},
  \bibinfo{author}{Shriram \surnamestart Krishnamurthi\surnameend} \&
  \bibinfo{author}{Matthias \surnamestart Felleisen\surnameend}
  (\bibinfo{year}{1997}): \emph{\bibinfo{title}{Dr{S}cheme: A pedagogic
  programming environment for scheme}}.
\newblock In \bibinfo{editor}{Hugh \surnamestart Glaser\surnameend},
  \bibinfo{editor}{Pieter \surnamestart Hartel\surnameend} \&
  \bibinfo{editor}{Herbert \surnamestart Kuchen\surnameend}, editors: {\sl
  \bibinfo{booktitle}{Programming Languages: Implementations, Logics, and
  Programs}}, {\sl \bibinfo{series}{Lecture Notes in Computer Science}}
  \bibinfo{volume}{1292}, \bibinfo{publisher}{Springer Berlin / Heidelberg},
  pp. \bibinfo{pages}{369--388}, \doi{10.1007/BFb0033856}.

\bibitemdeclare{book}{car}
\bibitem{car}
\bibinfo{author}{Matt \surnamestart Kaufmann\surnameend},
  \bibinfo{author}{J.~Strother \surnamestart Moore\surnameend} \&
  \bibinfo{author}{Panagiotis \surnamestart Manolios\surnameend}
  (\bibinfo{year}{2000}): \emph{\bibinfo{title}{Computer-Aided Reasoning: An
  Approach}}.
\newblock \bibinfo{publisher}{Kluwer Academic Publishers},
  \bibinfo{address}{Norwell, MA, USA}, \doi{10.1007/978-1-4615-4449-4}.

\bibitemdeclare{article}{page-se}
\bibitem{page-se}
\bibinfo{author}{Rex \surnamestart Page\surnameend} (\bibinfo{year}{2007}):
  \emph{\bibinfo{title}{Engineering Software Correctness}}.
\newblock {\sl \bibinfo{journal}{Journal of Functional Programming}}
  \bibinfo{volume}{17}, pp. \bibinfo{pages}{675--686},
  \doi{10.1017/S095679680700634X}.

\bibitemdeclare{inproceedings}{page-proptest}
\bibitem{page-proptest}
\bibinfo{author}{Rex \surnamestart Page\surnameend} (\bibinfo{year}{2011}):
  \emph{\bibinfo{title}{Property-Based Testing and Verification: a Catalog of
  Classroom Examples}}.
\newblock In: {\sl \bibinfo{booktitle}{Proceedings of the 2011 Symposium on
  Implemenation and Application of Functional Languages}},
  \bibinfo{address}{Lawrence, KS}, pp. \bibinfo{pages}{134--147},
  \doi{10.1007/978-3-642-34407-7\_9}.

\bibitemdeclare{inproceedings}{hcw}
\bibitem{hcw}
\bibinfo{author}{Rex \surnamestart Page\surnameend} \& \bibinfo{author}{Ruben
  \surnamestart Gamboa\surnameend} (\bibinfo{year}{2012}):
  \emph{\bibinfo{title}{How Computers Work: Computational Thinking for
  Everyone}}.
\newblock In: {\sl \bibinfo{booktitle}{Proceedings of the First International
  Workshop on Trends in Functional Programming in Education}},
  \bibinfo{address}{St Andrews, UK}, \doi{10.4204/EPTCS.106.1}.

\bibitemdeclare{inproceedings}{dracula}
\bibitem{dracula}
\bibinfo{author}{Dale \surnamestart Vaillancourt\surnameend},
  \bibinfo{author}{Rex \surnamestart Page\surnameend} \&
  \bibinfo{author}{Matthias \surnamestart Felleisen\surnameend}
  (\bibinfo{year}{2006}): \emph{\bibinfo{title}{{ACL2} in {D}r{S}cheme}}.
\newblock In: {\sl \bibinfo{booktitle}{Proceedings of the sixth international
  workshop on the ACL2 theorem prover and its applications}},
  \bibinfo{series}{ACL2 '06}, \bibinfo{publisher}{ACM}, \bibinfo{address}{New
  York, NY, USA}, pp. \bibinfo{pages}{107--116}, \doi{10.1145/1217975.1217999}.

\bibitemdeclare{article}{subjects}
\bibitem{subjects}
\bibinfo{author}{R.A. \surnamestart Virzi\surnameend} (\bibinfo{year}{1992}):
  \emph{\bibinfo{title}{Refining the test phase of usability evaluation: How
  many subjects is enough?}}
\newblock {\sl \bibinfo{journal}{Human Factors: The Journal of the Human
  Factors and Ergonomics Society}} \bibinfo{volume}{34}(\bibinfo{number}{4}),
  pp. \bibinfo{pages}{457--468}.

\end{thebibliography}
\end{document}